# Laser-Induced Skyrmion Writing and Erasing in an Ultrafast Cryo-Lorentz Transmission Electron Microscope

## Authors

G. Berruto[1†], I. Madan[1†], Y. Murooka[1,∥], G. M. Vanacore[1], E. Pomarico[1], J. Rajeswari[1], R. Lamb[2], P. Huang[1,3], A. J. Kruchkov[3], Y. Togawa[2,4,5], T. LaGrange[6], D. McGrouther[2], H. M. Rønnow[3] and F. Carbone[1*]

## Affiliations

[1]*Institute of Physics, LUMES, École Polytechnique Fédérale de Lausanne (EPFL), Lausanne CH-1015, Switzerland.*

[2]*Scottish Universities Physics Alliance, School of Physics and Astronomy, University of Glasgow, Glasgow G12 8QQ, United Kingdom;*

[3]*Institute of Physics, LQM, École Polytechnique Fédérale de Lausanne (EPFL), Lausanne CH-1015, Switzerland.*

[4] *Osaka Prefecture University, 1-2 Gakuencho, Sakai, Osaka 599-8570, Japan*

[5] *Chirality Research Center (CResCent), Hiroshima University, Higashi-Hiroshima, Hiroshima739-8526, Japan*

[6]*Interdisciplinary Centre for Electron Microscopy, École Polytechnique Fédérale de Lausanne (EPFL), 1015 Lausanne, Switzerland*

[†] These authors contributed equally to this work

[∥] Present address: *Ernst Ruska-Centre for Microscopy and Spectroscopy with Electrons* and *Peter Grünberg Institute, Forschungszentrum Jülich, 52425 Jülich, Germany*

[*] Corresponding author, e-mail: fabrizio.carbone@epfl.ch

**We demonstrate that light-induced heat pulses of different duration and energy can write skyrmions in a broad range of temperatures and magnetic field in FeGe. Using a combination of camera-rate and pump-probe cryo-Lorentz transmission electron microscopy, we directly resolve the spatiotemporal evolution of the magnetization ensuing optical excitation. The skyrmion lattice was found to maintain its structural properties during the laser-induced demagnetization, and its recovery to the initial state happened in the sub-µs to µs range, depending on the cooling rate of the system.**

Skyrmions are nanoscopic magnetic vortices with a nonzero topological charge. Creation, annihilation and motion of skyrmions in topological magnets offer interesting perspectives for spintronics and data storage devices[1,2]. While the former requires a simple and energy efficient way of manipulating the magnetic states, the latter aims at a fast and reliable way of writing/erasing and reading of individual skyrmions or skyrmion lattice segments in a confined volume. Furthermore, the study of the dynamics of the topological charge emergence introduces the concept of time into the physics of topological matter. Therefore, investigating the speed limit of skyrmion creation/annihilation is paramount for both fundamental and applied perspectives.

In general, the rate of skyrmions appearance as well as their persistence range in the host material's phase diagram depend on the balance between thermodynamic and topological stability[3]. Significant advances have been achieved in the control and manipulation of skyrmions in various hosting systems, ranging from insulating[4-8] to metallic[9-13], bulk and nanostructured helimagnets, to ultrathin heavy metals/ferromagnetic multilayered amorphous films [14-18]. To date, two approaches compatible with current technology have been demonstrated: the application of an electric field, which leads to the skyrmion lattice rotation in insulating $Cu_2OSeO_3$[4,7], and spin-transfer torque induced skyrmion motion at ultralow current flow in metallic systems[9,10,12,16]. As interesting alternatives, it has been shown that nanosecond magnetic field pulses can lead to the appearance and precession of skyrmions and skyrmion lattice segments[8,16,18], and it was demonstrated that collective modes of the skyrmion lattice were successfully induced by femtosecond (fs) laser pulses via the inverse Faraday effect, in simple words providing an effective magnetic field ultrashort pulse[6]. However, all of these observations were obtained via ensemble measurements and, to date, no spatially-resolved information on the dynamical response of skyrmions to time-varying stimuli and their intrinsic creation/annihilation speed is available. Such an information, besides addressing the fundamental question of how fast skyrmions can be created, is of pivotal importance for understanding the role of defects, edges and nanostructuring on the control of skyrmions. Furthermore, the creation or annihilation of skyrmions solely by optical pulses has remained elusive so far and would provide a unique handle in magneto-optical devices.

In a recent body of work, magnetic bubbles* were photogenerated in thin-films of ferrimagnetic rare earth-Fe-Co alloys[21,22,23,24]. Such magnetic textures are stabilized by uniaxial anisotropies and dipole-dipole interaction. The mechanism for their creation is based on the transient local heating above the Curie temperature[22,25,26] induced by a circularly polarized fs laser pulse[21,22,23]. Upon relaxation, the switched region defines the core of the magnetic bubble, and its size (≈0.3 µm or larger[23]) and shape are determined by the beam profile and intensity.

Skyrmions in chiral magnets are fundamentally different from the magnetic bubbles. They are stabilized by the competition between the magnetic exchange and the Dzyaloshinsky-Moriya (DM) interactions, and present a continuously whirling distribution of spins with a fixed chirality[2]. Such a spatial texture can be dramatically confined (down to few nm), and is intrinsically determined by the properties of the host material.

---

* Magnetic bubbles[19,20] are spin textures with a uniformly magnetized core separated from the oppositely magnetized surrounding by a domain wall. They are usually topologically trivial but can as well have non-trivial topological charges (chiral bubbles), as skyrmions. Unlike skyrmions, whose topological charge is defined by the sign of the Dzyaloshinsky-Moriya interaction, the topological charge of magnetic bubbles is not uniquely given by the material properties, although the magnetization textures can be engineered by nano-fabrication, for example tuning the dipole-dipole interaction via changes of a thin film thickness.

Here, we report the generation and dynamical evolution of skyrmions in the prototypical itinerant chiral magnet FeGe, initiated by laser-induced heat pulses. Contrary to the previously reported mechanisms[23,26,27], our approach allows the creation of skyrmions without transiently reaching the paramagnetic state. Moreover, their size is independent on the laser beam diameter and fluence. These experiments provide an example of the optically driven edge-injection of topological charges.

To determine the time needed for skyrmions to be created or erased by laser pulses, we performed a combination of in-situ cryo-Lorentz Transmission Electron Microscopy (cryo-LTEM) and nanosecond (ns) pump-probe cryo-LTEM[28,29] (Fig. 1a.). The experiments were carried out in our modified JEOL JEM2100 TEM[30]. In this instrument, in-situ cryo-LTEM can be performed in Fresnel configuration[31] at camera-rate temporal resolution (ms) using a continuous wave electron beam generated thermo-ionically, upon in-situ pulsed optical excitation of the specimen with tunable fs source (Supplementary Note I).

Nanosecond time-resolved stroboscopic experiments were carried out in pump-probe mode. The ns electron probe-pulses were photoemitted from a cathode exposed to a train of ultraviolet light pulses, obtained via the fourth harmonic generation of the fundamental emission of a Nd:YAG ns laser in nonlinear crystals. The delay with respect to a train of visible light (2.33 eV) pump pulses illuminating the TEM sample was controlled via a digital delay generator. The laser spot size was kept at least 30 times larger than the sample lateral dimension providing homogeneous illumination conditions.

The 60 nm thick FeGe sample was inserted in a liquid nitrogen cryogenic holder enabling temperature control between 100 and 300 K, and the external perpendicular magnetic field was controlled by variable objective lens excitation. The phase diagram of our nano-slab was determined by performing cryo-LTEM experiments at different temperatures and under different applied magnetic fields. A phase diagram sketch based on these experiments is depicted in Fig. 1b.

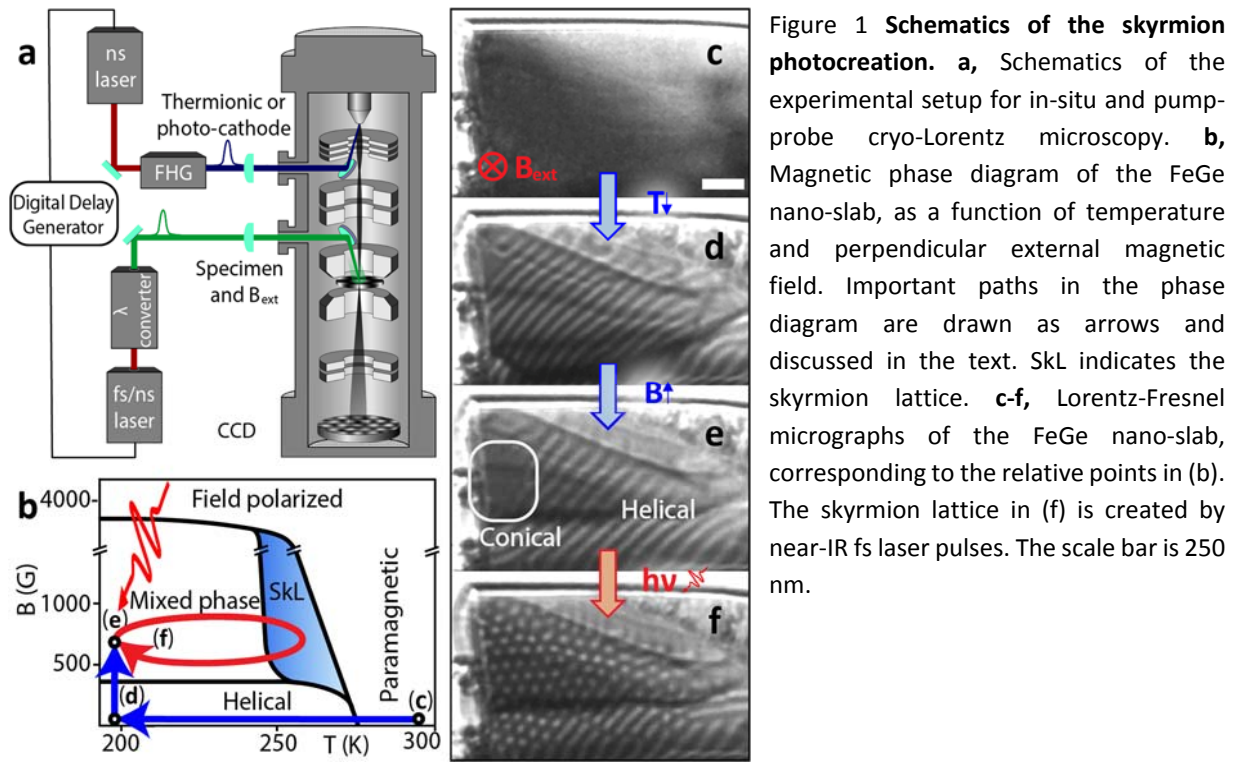

Figure 1 **Schematics of the skyrmion photocreation. a,** Schematics of the experimental setup for in-situ and pump-probe cryo-Lorentz microscopy. **b,** Magnetic phase diagram of the FeGe nano-slab, as a function of temperature and perpendicular external magnetic field. Important paths in the phase diagram are drawn as arrows and discussed in the text. SkL indicates the skyrmion lattice. **c-f,** Lorentz-Fresnel micrographs of the FeGe nano-slab, corresponding to the relative points in (b). The skyrmion lattice in (f) is created by near-IR fs laser pulses. The scale bar is 250 nm.

The static temperature-dependent characterization (Fig. 1c-e) of the magnetization pattern and the skyrmion photocreation (Fig. 1f) were observed in cryo-LTEM in Fresnel mode, using a defocus length of ∼380 µm. In the paramagnetic state, LTEM shows no magnetic contrast (Fig. 1c). When cooled down below the Curie temperature, $T_C \approx 278$ K, in the minimum magnetic field achievable in our set-up, 170 G, the material transforms to the helical phase, which is visible in LTEM as stripes[32], see Fig. 1d. At higher magnetic field and low temperature, the conical phase, which is indistinguishable from the field-polarized state in LTEM, is expected to be found based on previous experiments on wedged samples[33] and theoretical considerations[34]. In our sample, we observe either a coexistence of conical and helical phase, or the skyrmion phase, depending on the path followed in the phase diagram. The skyrmion lattice (SkL) is a ground state in the narrow region of phase space visible in Fig. 1b, above a characteristic field-dependent temperature $T_{SkL\uparrow}$ irrespective of the initial magnetic texture, and, once formed, was observed to persist upon field cooling down to the lowest temperature that we could reach in our set-up, 100 K. In other words, the magnetic state of the sample in the "Mixed phase" region of the phase diagram strongly depends on the field-dependent cooling/heating history, highlighting the hysteretic nature of the magnetic phases in FeGe. The initial state for the photocreation experiments was prepared via the path c-d-e depicted in Fig. 1b, so that the sample was predominantly in the conical/helical phase (Fig. 1e), with no skyrmion present.

From this region of the phase diagram, upon the in-situ application of laser pulses, it is possible to write skyrmions in the material, see Fig. 1f and Supporting Movies 1 and 2. Skyrmions can then be fully erased by walking into the helical or field polarized phases and back, just by exciting the external magnetic field, at constant temperature. Large skyrmion lattice segments can be also erased by high fluence optical pulses, see discussion below. We found that starting from 233 K, in a magnetic field of 400 G, a single 1.55 eV, 60 fs optical pulse of as little as 2 mJ/cm² is necessary to create a cluster of 10 to 14 skyrmions. Given the typical skyrmion diameter of 70 nm, and the knowledge of FeGe optical absorption, a rough estimate of the optical energy expense is $8 \times 10^{-15}$ J/skyrmion, offering interesting perspectives for energy efficient data storage[35,36].

To determine the mechanism for the light-induced skyrmion writing, we measured a fluence-dependence series and identified the energy per pulse necessary to create skyrmions starting from the different positions in the phase diagram, indicated by open circles in Fig. 2a. We estimated the threshold value by observing the magnetic pattern and gradually increasing the laser illumination fluence. The critical fluence is taken as the value to create well defined skyrmions, and its error bar (approximately 18%) is estimated by repeating the same experiment three times. The distribution of threshold values obtained for the different sample conditions is plotted as a color map in Fig. 2a. In this graph, we also plot both the warming (red dots) and cooling (dashed red line) skyrmion stability regions of the material measured in absence of any photoexcitation. Above $T_{SkL\uparrow}$ the skyrmion lattice forms thermally without external stimuli.

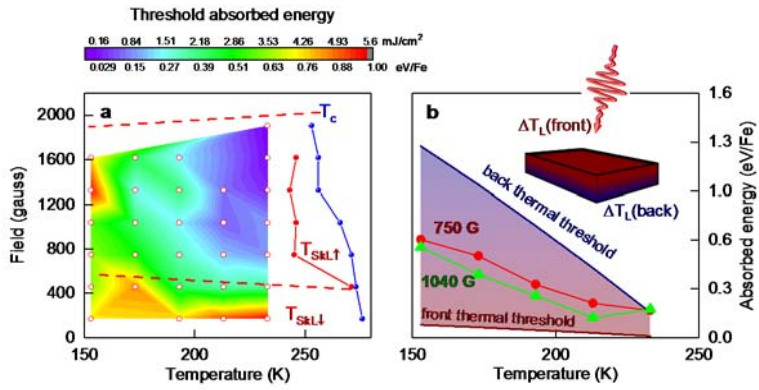

Figure 2 **Skyrmion photocreation phase diagram. a,** the color map encodes the skyrmion photocreation energy threshold as a function of temperature and magnetic field. 60 fs long, linearly polarized 1.55 eV pulses were used. For comparison, the temperature at which skyrmions form upon field-warming ($T_{SkL\uparrow}$) and the critical temperature for the skyrmion to paramagnetic transition ($T_C$) were measured experimentally and are shown as full symbols. The region of stability of the skyrmions after field cooling is enclosed between the red dashed lines $T_{SkL\downarrow}$. **b,** temperature dependence of the threshold at selected magnetic fields. For comparison, the fluence values required to warm-up the lattice to $T_{SkL\uparrow}$ at the front (top) and back (bottom) surfaces of the sample are shown in red and blue respectively, for $T_{SkL\uparrow}$=243 K.

As visible, at lower temperatures a larger energy is required to create skyrmions, suggesting that the light-induced temperature rise in the material is responsible for their appearance. To test this hypothesis, we calculated the optical energy needed to cross $T_{SkL\uparrow}$ as a function of the starting temperature for both the top and bottom surfaces of the sample (Supplementary Note I). These values are represented as solid red and blue lines in Fig. 2b, respectively. The measured absorbed laser energy threshold values as a function of the starting temperature for a magnetic field of B=750 G and B=1040 G are also displayed as red and green symbols, respectively. The experimental laser energy absorbed by the sample required to create skyrmions largely exceeds the theoretical value for the top surface and lies somewhat below the value for the bottom surface, suggesting that sufficiently thick portions of the slab surpass the critical temperature to induce the skyrmions formation.

Importantly, the electronic and lattice temperature follow a different dynamical evolution upon excitation by fast laser pulses. The light predominantly couples to the electronic subsystem resulting in a transient increase of the electronic temperature. The equilibration of the electronic and structural subsystems is mediated by the electron-phonon coupling and can be described by a two-temperature model (2TM)[37], see Supplementary Note I. Typically, the shorter the light pulses, the higher the electronic temperature, and the larger its difference to the lattice temperature at the early stage of the dynamics. Such a behavior is depicted in Fig. 3a, where the temporal evolution of the electronic and lattice temperature was obtained via the 2TM as a function of pulse duration.

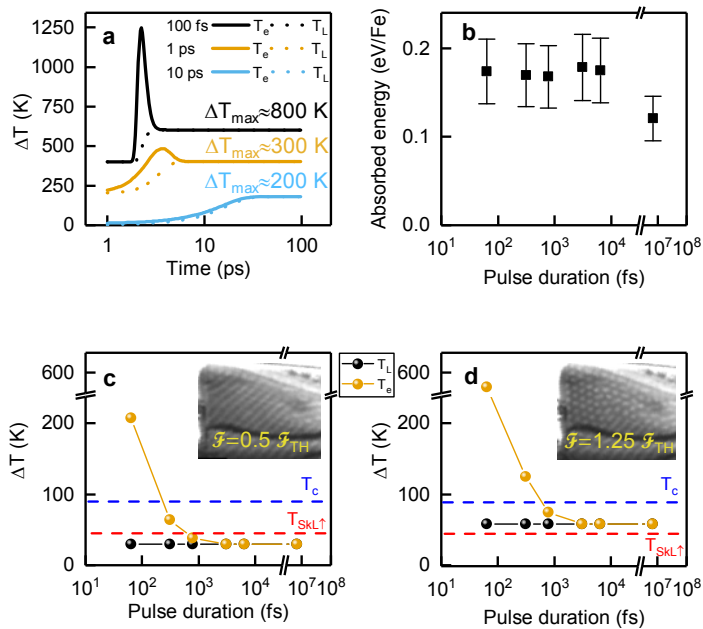

Figure 3 **Electron and lattice temperature evolution. a,** Two-temperature model calculation of the electronic and lattice temperature evolution for three pulse durations, 100 fs, 1 ps, and 10 ps at the surface of the sample (solid and dashed for electronic and lattice temperatures, respectively). **b,** Absorbed energy threshold for photocreation of skyrmions as a function of pulse duration. **c, d,** Peak variation of the depth-averaged electronic and lattice temperatures as a function of pulse duration at one half and just above the threshold fluence values. Red and blue dashed lines indicate temperature variation required for thermal creation of skyrmions and critical temperature $T_C$ upon field-warming. The insets show the cryo-LTEM images taken upon photoexcitation of the sample at 233 K with 1.55 eV, 60 fs laser pulses.

To disentangle the role of these two subsystems, we performed a series of skyrmions photocreation threshold measurements as a function of the pulse duration, shown in Fig. 3b. It is evident that the threshold is independent of the laser pulse duration, suggesting that the photoinduced electronic temperature jump does not play a role in the formation of the skyrmions. To corroborate this idea, we show the photo-induced electronic and lattice temperature jump for different laser pulse durations above and below the threshold for photo-generating skyrmions, Fig. 3c and 3d. When only the electronic temperature far exceeds the critical value $T_{SkL\uparrow}$, no effect is observed on the magnetic pattern; conversely, if the lattice temperature is above $T_{SkL\uparrow}$, skyrmions are observed in the sample. The threshold for the photo-generation of skyrmions is also independent on the photon energy between 0.8 and 1.55 eV (see Supplementary Fig. 1). This is expected for a thermally-induced phenomenon considering the rather featureless optical conductivity of FeGe in such an energy range[38].

The fact that a very large electronic temperature does not result in observable changes of the system's magnetization, together with the insensitivity of the threshold to the driving pulses duration and wavelength, implies that the mechanism for creating skyrmions is a slow process. In the single-pulse writing experiments, the phase diagram after the excitation results changed into the cooling phase diagram, see Fig. 4a red arrow, following an evolution similar to supercooling[39,40]. The sudden quench of thermal fluctuations ($\sim$K/µs) that takes place after the initial temperature jump freezes skyrmions in regions of the phase diagram where they were not present before the arrival of the laser pulse. This is consistent with previous reports showing that fast cooling rates, in the range of K/s, can homogeneously freeze-in skyrmions in different regions of the phase diagram[41]. Importantly, in our case the skyrmion creation occurs in the limited region of the sample either close to an edge or at the interface between magnetic phases, whilst the majority of the sample conserves its magnetic state (see

Supporting Movies 1 and 2). This is different from the slower equilibrium-cooling experiments when most of the sample converts into the skyrmion lattice, and showcases the role of the cooling rate in tipping the balance between topological protection and thermal fluctuations. In fact, the emergence of the skyrmions is expected to happen at the edges of the sample where such a balance is altered by the breaking of topological protection. This has been experimentally observed upon magnetic field variation[42] and rationalized theoretically for the pulsed suppression of the magnetization[43].

To test the reaction speed of the sample's magnetization to a sudden temperature jump, we performed a ns pump-probe experiment in proximity of the skyrmion to paramagnetic phase transition, Fig. 4a orange arrow. A 20 kHz train of 25 ns, 2.33 eV laser pulses with variable fluence was used to perturb the skyrmion lattice close to the critical temperature $T_C$. The evolution of the skyrmion lattice was monitored via a synchronized train of 30 ns-electron pulses, and the time delay was varied electronically, see Supplementary Note I. We have chosen to study stroboscopically this transition because it is fully reversible when crossed at the slow rate of $<$ K/s , both heating and cooling, and does not involve any metastable state (see Supplementary Fig. 3). The FeGe slab was kept at 257 K and 950 G.

A portion of a typical Lorentz micrograph of the skyrmion lattice recorded with photoelectrons is depicted in Fig. 4b, together with its spatial Fourier Transform in the inset. The temporal evolution of the skyrmion lattice constant, orientation angle, and magnetization strength for two fluences (approximately 1.5 and 5 mJ/cm$^2$) are plotted in Fig. 4c top and bottom panels, respectively. We found no significant modification to either the skyrmion lattice constant or its orientation angle. As the skyrmion lattice constant is given by the ratio between the Dzyaloshinsky-Moriya D and the exchange constant J ($d = 4\pi J/|D|$), and the two are independent physical quantities, it is reasonable to conclude that neither D nor J are affected by the laser pulses on the ns time scale. Contrary, the intensity of the magnetic contrast, estimated as the intensity of the magnetic Bragg peak obtained by Fourier transforming the real-space image, is found to decrease by 6 and 49 percent for 1.5 mJ/cm$^2$ and 5 mJ/cm$^2$ fluence, respectively. The time scale for the recovery of the magnetization depends on the fluence and reaches ~9 μs for the strongest photoexcitation tested (5 mJ/cm$^2$), confirming the idea that the magnetization dynamics is governed by the heat diffusion.

We can now further understand the writing process. After the photon absorption, the energy is transferred to the lattice on the few-ps time scale. The consequent excitation of magnetic fluctuations favors the skyrmion creation at the sample edges and magnetic phase boundaries as long as T>$T_{SkL\uparrow}$. After ~1 μs, the system supercools down into the region of the phase diagram (T<$T_{SkL\uparrow}$) where magnetization does not further evolve. Since we only observe the formation of up to 14 skyrmions with each pulse and the optically written skyrmion lattice is highly disordered (compare Fig. 1f and 4b), we can claim that few μs are not sufficient for establishing long-range skyrmionic order, and skyrmions are formed as individual self-assembling entities.

To corroborate this idea, we also performed an experiment where a single high-fluence pulse ($> 2.5 \times \mathcal{F}_{TH}$) was applied to the sample in which the skyrmion lattice was already photoinduced (Fig. S2 b). At this photoexcitation level, the sample transient temperature exceeds $T_c$, resulting in its complete demagnetization. As a consequence, a large portion of the skyrmion lattice disappears, and only few skyrmions reform during cooling close to the edge of the specimen, see Fig. S2 c. Importantly, the skyrmion cluster size does not grow with the number of pulses and its position and shape differ after each individual laser pulse,

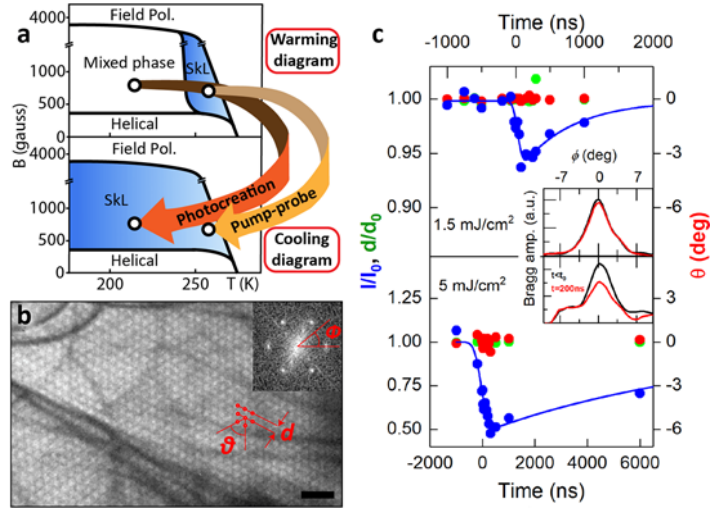

Figure 4 **Ns pump-probe Lorentz TEM on skyrmions. a,** Schematic evolution of the system across the phase diagram during the laser-writing (photocreation) and pump-probe experiments. **b,** Portion of the skyrmion lattice micrograph recorded with 25 ns photoelectron pulses and its Fourier transform, inset. Scale bar is 250 nm. **c,** Pump-probe temporal dynamics of the skyrmion lattice parameters: skyrmion-skyrmion distance ($d/d_0$, with $d_0$ the distance at equilibrium, green symbols – left axis), orientation angle (red symbols – right axis) and magnetization strength ($I/I_0$, where $I_0$ is the magnetic Bragg peak intensity at equilibrium, blue symbols – left axis) at approximately 1.5 and 5 mJ/cm² fluence, top and bottom panel respectively. The solid blue curves are fits to the magnetization strength via a finite rise-time exponential relaxation function, yielding recovery time constants of $0.75 \pm 0.3\ \mu s$ at 1.5 mJ/cm², and $9.3 \pm 4.4\ \mu s$ at 5 mJ/cm². The insets show the magnetic Bragg peak profile before the arrival of the pump pulse (black) and after 200 ns (red). The curves were obtained averaging over all six Bragg peaks after having verified that all have the same dynamics, ruling out distortions of the skyrmion lattice.

confirming that each time the skyrmion formation starts from the (transient) paramagnetic state.

In summary, we have shown that skyrmions can be written in a FeGe nano-slab via illumination with optical pulses of different colors and duration. Our space-time resolved observations highlighted that the light-induced formation of skyrmions takes place at the boundaries between magnetic domains and in proximity of the sample edges. The writing and erasing speed of the skyrmions was found to be governed by the cooling rate following the laser-induced temperature jump. In the present case, the speed limitation is a direct consequence of the heat diffusion rate (K/μs) in our nano-fabricated slab.

Based on our results, we suggest that increasing the cooling rate up to K/ps by connecting the sample to a heat-bath should result in the efficient skyrmion optical erasing by strong ps pulses. The writing process would still be done by long ns pulses, whose pulse duration determines the thermalization time. The control of the skyrmion cluster size could be attained by varying the duration of the writing pulse.

Furthermore, the modulation of the topological landscape via nanofabrication would also allow to tune the creation/annihilation speed of the skyrmions and their stability against external perturbations. A different behavior may be expected in non-metallic skyrmion hosts where the optical absorption can exhibit large energy gaps and light-induced excitations can have longer life-times. In such systems, also electronic effects may influence the fast dynamics of the skyrmions.

**Acknowledgments**

The LUMES laboratory acknowledges support from the NCCR MUST, Sinergia CDSII5-171003 Nanoskyrmionics and an Ambizione grant of the Swiss National Science Foundation. G. B. acknowledges financial support from the Swiss National Science Foundation (SNSF) through the grant 200021_159219/1. E. P. acknowledges financial support from the SNSF through an Advanced Postdoc Mobility Grant (P300P2_158473). HMR acknowledges support from the SNSF through the grant 200020_166298. The authors would like to acknowledge M. Miyagawa, Y. Kousaka, T. Koyama, J. Akimitsu, K. Inoue for the preparation and characterization of the bulk FeGe crystal, D. Demurtas, M. Cantoni and D. Alexander from CIME for technical support with the cryogenic holder, A. Rosch, and A. Bogdanov, for insightful discussions, T.T.L. Lummen and A. Pogrebna for help with the set-up and the experiments, respectively.

# Supplementary Information

## Supplementary Note I (Technical details of the experiments and simulations)

**Experiments:**

For the photocreation experiments, a Ti:Sapphire 1.55 eV laser in combination with an OPA (0.8 – 1.1 eV) was used to photoexcite the specimen. Wavelength dependent measurements are reported in the Supplementary figure 1. For the threshold determination, a 20 Hz pulse train was used to suppress CW heating effects. The reported values are averages over three measurements, and the associated error bar is their statistical spread. Single pulse experiments were conducted employing a mechanical shutter. The value of the threshold (expressed in eV/Fe site) was obtained from the incident threshold laser fluence knowing the penetration depth and reflectivity of FeGe from Ref. [1], which are about 23 nm and 57%, respectively. The optical energy expense to create a skyrmion was deduced from the absorbed optical fluence and the knowledge of the spot size of 80 µm and the average area occupied by a skyrmion (approximately 1400 nm$^2$).

For the ns pump-probe experiment, the second harmonic (2.33 eV) of a Nd:YAG laser was used to photoexcite the specimen, while the 4th harmonic (4.66 eV) of a second Nd:YAG laser, obtained through Fourth Harmonic Generation in nonlinear crystals, was used to generate photoelectrons from a flat Ta disk cathode. The delay was set by a digital delay generator and monitored by an oscilloscope.

**Sample:**

Thin lamellas were prepared by Ga ions FIB milling from a bulk FeGe crystal, employing FEI 600 Nanolab dual beam system. The lamella thickness was estimated from the percentage of inelastically scattered electrons in-situ, and the crystallinity of the specimen checked also in-situ by selected area electron diffraction.

**Simulations and data analysis:**

To calculate the temperature evolution two models were used. For the sub-10 ns dynamics we used the two-temperature model by Allen[2]. The thermal constants were taken from Ref. [3] while the electron-phonon relaxation constant was estimated from Ref. [1]. On the longer time-scales we have solved the one-dimensional heat diffusion equation on a length-scale of 40 times the sample size (real size of 500 nm was used). Having the laser spot much larger than the sample size, it was modeled by a homogeneous 20 ns long heat source. The Wiedemann-Franz law was used to estimate the heat conductivity from electrical transport data[4]. The quench rate was estimated from this solution at $T_c$, and it is of the order of K/µs.

The pump-probe data were fitted with a finite error-function rise-time exponential relaxation function. While the heat diffusion fit is more appropriate, it is not feasible to analytically solve the equation accounting for the actual shape of the sample and the heat sink. The exponential fit gives a reliable estimate of the initial remagnetization time scale.

# Supplementary figure 1

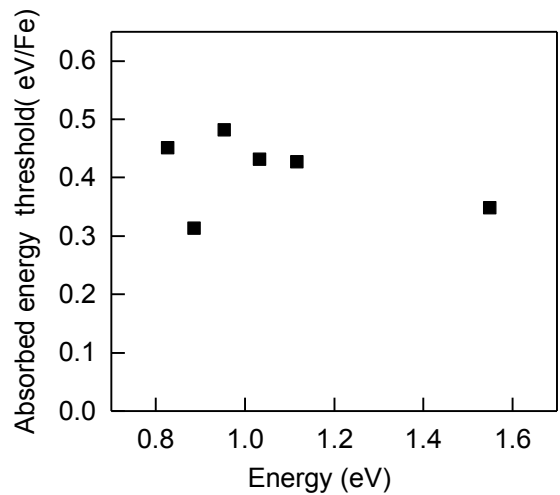

S.Fig 1 **Dependence of the SkL photocreation threshold on writing laser photon energy.** Measurements conducted at 213 K and 950 G. The absence of electronic resonances favors a dominant role of the lattice temperature in the photocreation process.

# Supplementary figure 2

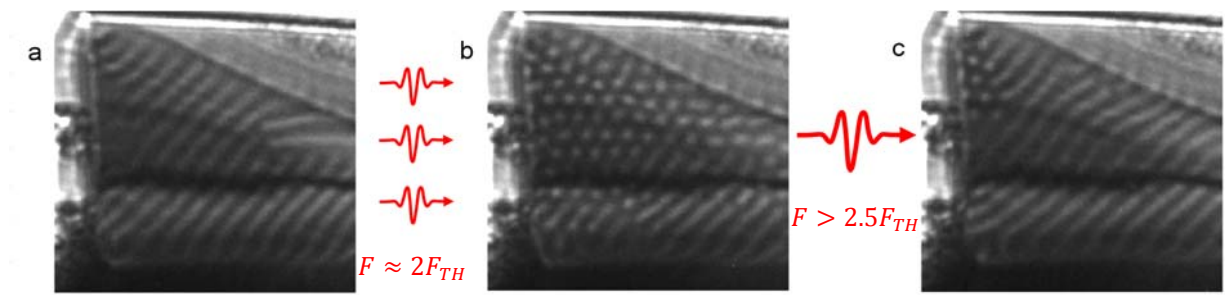

S.Fig 2 **Effect of high power pulses on skyrmion creation: optical annihilation of the skyrmion lattice. a,** Initial state at 213 K and 950 G. **b,** Skyrmion pattern created by a train of pulses with $F \approx 2F_{TH}$. **c,** Subsequent illumination with $F > 2.5\ F_{TH}$: only few skyrmions are present independently on the number of pulses.

# Supplementary figure 3

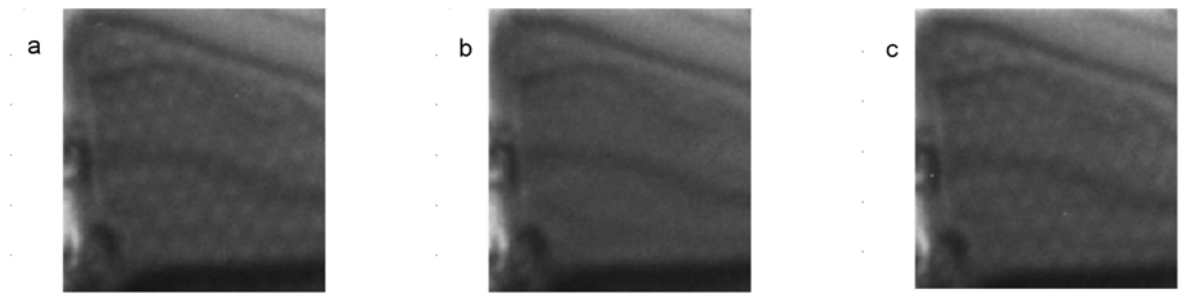

S.Fig 3 **Reversibility of the transition into the SkL phase.** Micrograph of the skyrmion phase 5 degrees below Tc **(a)**, at Tc **(b)**, and cooled again to the initial temperature **(c)**.

## Supporting Movie 1

Movie of the single-pulse skyrmion photocreation. Each frame is a 0.5 s integrated Lorentz-Fresnel micrograph taken at T=213 K and B=700 G. The magnetic state of the specimen remains stable until a light pulse is shined on it. After the skyrmion creation, whose transient state is not resolved with our camera-rate acquisition, the sample changes magnetization, which is stable until a following pulse impinges the sample.

## Supporting Movie 2

Movie of the single-pulse skyrmion photocreation. Each frame is a 0.5 s integrated Lorentz-Fresnel micrograph taken at T=213 K and B=700 G. Unlike in the Supporting Movie 1, between any consecutive frames the specimen is illuminated by one laser pulse.

# SI References